\pdfoutput=1

\documentclass[sigconf]{acmart}
\AtBeginDocument{%
  \providecommand\BibTeX{{%
    \normalfont B\kern-0.5em{\scshape i\kern-0.25em b}\kern-0.8em\TeX}}}
\usepackage{algorithm} 
\usepackage{algpseudocode} 
\setcopyright{acmcopyright}
\copyrightyear{2023}
\acmYear{2023}
\acmDOI{XXXXXXX.XXXXXXX}

\acmConference[Conference acronym 'XX]{Make sure to enter the correct
  conference title from your rights confirmation email}
\acmBooktitle{Unknown Conference, USA}

\newcommand{\arya}[1]{{\color{magenta}(Arya: #1)}}

\begin{document}

\title{Fire and Smoke Digital Twin -- A computational framework for modeling fire incident outcomes}

\author{Junfeng Jiao}
\authornote{Both authors contributed equally to this research.}
\email{jjiao@austin.utexas.edu}
\orcid{1234-5678-9012}
\author{Ryan Hardesty Lewis}
\authornotemark[1]
\email{rhl@utexas.edu}
\affiliation{%
  \institution{University of Texas at Austin}
  \streetaddress{P.O. Box 1212}
  \city{Austin}
  \state{Texas}
  \country{USA}
  \postcode{43017-6221}
}

\author{Kijin Seong}
\affiliation{%
  \institution{University of Texas at Austin}
  \streetaddress{1 Th{\o}rv{\"a}ld Circle}
  \city{Austin}
  \state{Texas}
  \country{USA}}
\email{kijin.seong@austin.utexas.edu}

\author{Arya Farahi}
\affiliation{%
  \institution{University of Texas at Austin}
  \city{Austin}
  \state{Texas}
  \country{USA}
}\email{arya.farahi@austin.utexas.edu}

\author{Paul Navratil}
\affiliation{%
 \institution{Texas Advanced Computing Center}
 \streetaddress{Rono-Hills}
 \city{Austin}
 \state{Texas}
 \country{USA}}

\author{Nate Casebeer}
\affiliation{%
  \institution{Austin Fire Department}
  \streetaddress{30 Shuangqing Rd}
  \city{Austin}
  \state{Texas}
  \country{USA}}

\author{Dev Niyogi}
\affiliation{%
  \institution{University of Texas at Austin}
  \streetaddress{8600 Datapoint Drive}
  \city{Austin}
  \state{Texas}
  \country{USA}
  \postcode{78229}}
\email{dev.niyogi@jsg.utexas.edu}

\renewcommand{\shortauthors}{Jiao and Lewis, et al.}

\begin{abstract}
Fires and burning are the chief causes of particulate matter (PM2.5), a key measurement of air quality in communities and cities worldwide. This work develops a live fire tracking platform to show active reported fires from over twenty cities in the U.S., as well as predict their smoke paths and impacts on the air quality of regions within their range. Specifically, our close to real-time tracking and predictions culminates in a digital twin to protect public health and inform the public of fire and air quality risk. This tool tracks fire incidents in real-time, utilizes the 3D building footprints of Austin to simulate smoke outputs, and predicts fire incident smoke falloffs within the complex city environment. Results from this study include a complete fire and smoke digital twin model for Austin. We work in cooperation with the City of Austin Fire Department to ensure the accuracy of our forecast and also show that air quality sensor density within our cities cannot validate urban fire presence. We additionally release code and methodology to replicate these results for any city in the world. This work paves the path for similar digital twin models to be developed and deployed to better protect the health and safety of citizens. 
\end{abstract}





\maketitle

\section{Introduction}

A digital twin is often described as ``a digital representation of a physical item or assembly using integrated simulations and service data'' \citep{Vrabi2018DigitalTU}. In the pursuit of an eventual all-encompassing ``Smart City,'' or a digital twin of an entire city's infrastructure, there have been numerous research ventures into digitizing city information, such as flood modeling, traffic prediction, emergency evacuation routing, among other applications \citep{10.1145/3453172,ong2022openems}. The United Nations predicts that most of the world's population currently lives within cities. Coupled with the fact that the probability of natural disasters will only increase as climate change progresses \citep{inbook}, emergency management and public services must be scaled to meet this future. In particular, active monitoring systems be developed to track, warn, and even predict disastrous events \citep{predicteq,predictfloods}. A large number of disastrous events are dangerous because resulting air pollutants, like asbestos, pesticides, and cyanide, find themselves airborne and transported downwind via smoke to affect nearby populations \citep{Asbestos,wildfirehealth,firepesticides}. As wildfires and other fire events grow in magnitude, it's hard not to realize that urban fires are \textit{not} normal, or at least they used to be much less frequent. 

Smoke impacts from fire events have dangerous consequences \citep{urbanfiresmoke}. Air pollution has been found to create an increased rate of morbidity and mortality, both causing and accelerating the progress of many diseases and conditions \citep{airpol}. Moreover, 80\%-90\% of fire-related deaths are attributed to smoke inhalation \citep{smokeinhalation}, and even a sustained small change in air quality can increase the mortality rate of some diseases by twentyfold \citep{covid}. To meet this challenge of often-occurring yet widely ignored health hazards from local fires, we develop \textsc{FireCOM},  a ``digital twin'' model that focuses on urban fires with real-time tracking and prediction of their smoke falloff, for community and neighborhood level awareness. This tool is designed to help the city officials in planning stage and mitigating the threat of urban fires and their aftermaths.

Fires detection and tracking is usually performed using satellite data. The fire detection models often relies on abnormal heat signatures and bright spots on the Earth's surface \citep{https://doi.org/10.5067/firms/modis/mcd14dl.nrt.0061, Dozier}. This type of tracking is helpful for wildfires, which are so large that they appear on satellite data. Their distinct and high signal-to-noise heat signal makes it easy to rely on satellite data. These signals for the fire incidents in an urban environment, however, are low signal-to-noise and often indistinguishable from the white noise, which makes detection difficult and leads to many false positives due to other source of noise, such as sun-glint off metallic structures, or false negatives \citep{Roy2006TheUO}. False negatives are more common, though. Often there are no detections, as not all urban fires are large enough for spatial satellite resolution and often remain covered by trees, buildings, and other obstructions. An alternative route would be to use 911 calls, but not much is done with this data. Many U.S. cities provide live pages to the public of emergency incidents as they are reported. These reports can be accurately pinpointed to a street address and the time lags are insignificant. \textsc{FireCOM} utilizes this information to track urban fires.

Tracking fires in real-time is not the only challenge. It is more important to model and predict the impact of urban fire incidents, specifically particulate matter exposure, such that the decision-makers can react rapidly as the situation unfolds. The existing fire and smoke studies mainly focus on wildfires \citep{HighResolutionSmokeForecasting,sfire,wildfireplan}. Wildfire impact is often larger -- spans hundreds and thousands of acres -- making these models less practical for vulnerable urban residents avoiding exposure to smoke. While urban fire studies are not non-existing \citep{urbanfiremodel,urbanfirerisk}, not much work is done in this area. This is despite the fact that fire incidents are more widespread and common in urban communities. \textsc{FireCOM} fills this gap.

\textsc{FireCOM} gathers real-time fire and building footprint data, and combines this information with weather data to accurately predict smoke dispersion within one, two, and three hours following a fire incident. This model not only helps to warn individuals about deteriorating air quality in their vicinity, but also assists fire departments and decision-makers in coordinating their efforts to mitigate the fire and its smoke-related aftermath.

The primary users of FireCOM include urban residents, emergency responders, and decision-makers involved in fire management and public health. By serving as an early warning system, \textsc{FireCOM} aims to improve health conditions and facilitate a more effective response to urban fires. The tool's accuracy and effectiveness are validated through data from thousands of air quality sensors and through collaborative case studies with the Austin Fire Department, ensuring its reliability and practical applicability in predicting urban fire impacts on air quality.

\section{Related Work}

The toxicity of wildfire smoke and urban smoke is a significant health hazard \citep{alarie_2002,firepesticides,Asbestos}. Carbon monoxide is often the leading cause of deaths in fires, but burning synthetic materials, such as those in urban environments, produces cyanide in fires that causes up to 50\% of fire deaths \citep{alarie_2002}. Building structures are often made of inorganic materials, implying that fumes from burning structures could be more harmful than forest fires. Fire incidents in urban environments, where the majority of fires are structure fires \citep{structurefireanalysis} compounded with high population density compared to rural areas, present an acute and immediate danger to vulnerable communities. Hence, real-time tracking of smoke falloff and mitigating its harms can save lives. Actively tracking, predicting, and managing smoke is usually focused on wildfires, with examples including NOAA's AirNow forecast \citep{airnow} and the US Forest Service's BlueSky \citep{bluesky}. However, little work has been done in predicted smoke from urban fires, with existing models focusing on evaluating risk and fire spread \citep{Li2013,urbanfiremodel,urbanfirerisk}. To the best of our knowledge, our model provides the first city-wide tracking and prediction of urban smoke.

\begin{figure}[h]
  \centering
  \includegraphics[width=\linewidth]{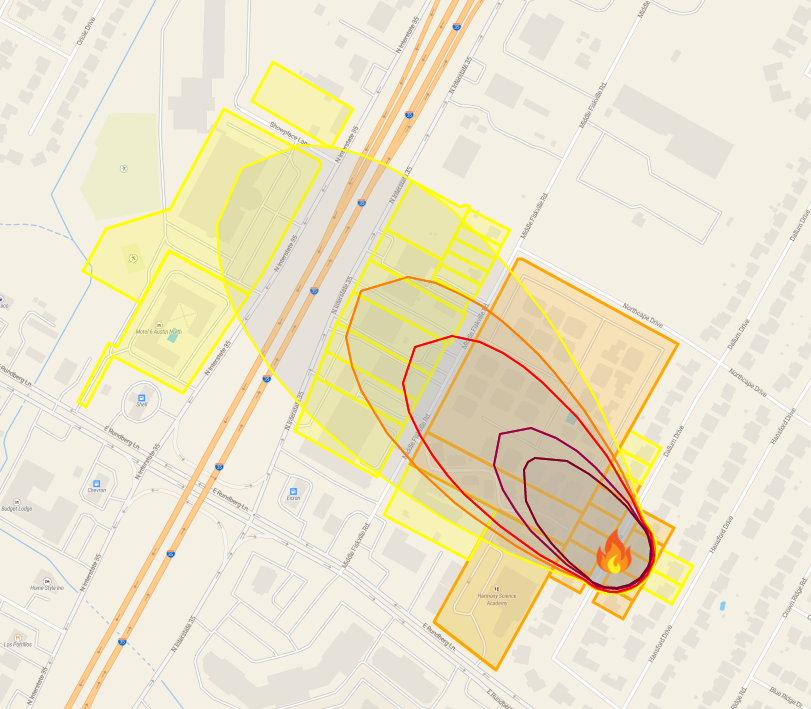}
  \caption{A VSmoke smoke simulation in-browser, with residences affected also highlighted according to smoke impact.}
  \Description{A VSmoke smoke simulation in-browser, with residences affected also highlighted according to smoke impact. \arya{Please never include a figure not referenced in the paper. Figures are there to guide and provide support for the arguments and storytelling. The reader cannot understand what we have to say about this figure, how it is related to this work, and where in the text they should look at this figure, and how it should be interpreted. }}
  \label{figure1}
\end{figure}

Prediction of the fire smoke diffusion process requires running computationally expensive fluid dynamics numerical simulations, but results in near-realistic outcomes.   
For instance, \citet{ZHAO2022101755} used their smoke simulation to recommend changes to typical subway carriages and proposed guidelines for evacuation protocols. Their results achieved the same performance as an empirical miniature tunnel study done by \citet{LI2012690} almost a decade earlier. Fluid dynamic simulations have become a tool of engineering, design, and prediction with verifiable accuracy \citep{fluidsimprogress}. However, for these simulations to be useful outside of a scientific audience, visualization is key. We combine domain-specific visualizations of both fluid-simulated smoke and a 3D city superimposed in order to make our forecasts publicly available, through a web dashboard, to urban residents. Alongside the help of real-time air sensors, the combination of these various tools allows us to model smoke dynamics on a city-wide scale. Figure \ref{figure1} shows the early ability of our approximate VSmoke forecast in determining the land uses an urban fire affects and the degree to which they will be affected.

\begin{figure*}[h]
  \centering
  \includegraphics[width=\linewidth]{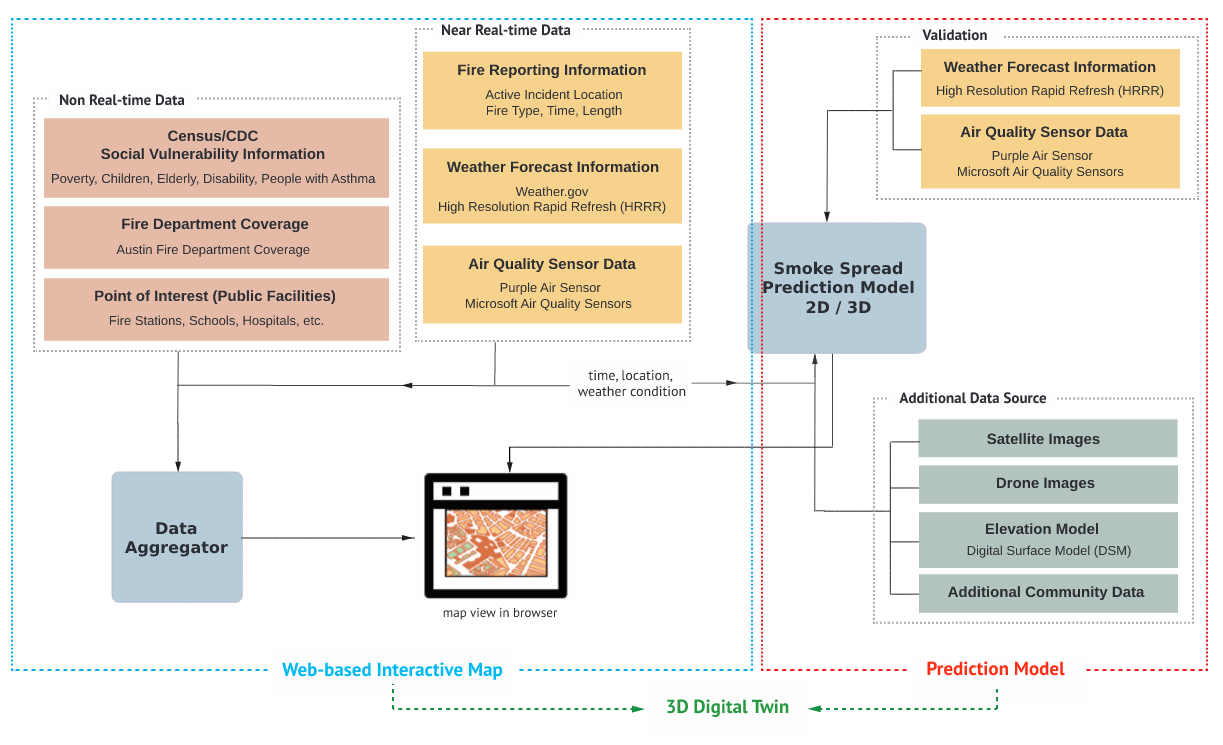}
  \caption{A diagram of the urban smoke prediction workflow.}
  \Description{A diagram showcasing the urban fire workflow.}
  \label{figure2}
\end{figure*}

\section{\textsc{FireCOM} a digital twin architecture for active fires}

\textsc{FireCOM} is a digital twin architecture designed for real-time monitoring and prediction of active fires and their impact on urban environments. Our data comprises three main components: (1) fire incidents, (2) weather data, and (3) air quality data. We obtain real-time fire incident locations from the Austin Fire Department, weather and wind conditions surrounding each fire from the National Weather Service, and air quality data from nearby outdoor air quality sensors provided by PurpleAir's sensors\footnote{\url{https://api.purpleair.com/}}.

Blender, a powerful open-source 3D creation suite, is chosen as the research platform due to its easy integration with Python, extensibility, and compatibility with GIS tools and accurate fluid simulations. Although Blender may not be primarily designed for real-time work, it offers precise models and simulations, essential for our research. We integrate OpenTopography's topographic map of Austin and OpenStreetMap's 3D city structures to create the digital twin's base map.

To improve the accuracy of our simulations, we incorporate real-world infrastructure, including building and street classifications, such as schools, hospitals, and speed limits. This enables our model to identify areas at higher risk from reduced air quality.

Our fire model leverages live information from local fire departments, weather services, and air quality sensors to predict smoke paths and monitor fire incidents. We map reported fire incidents using geographical coordinates, and then query the National Weather Service for recent weather and wind information at each location. Subsequently, we generate predicted smoke paths for each fire and validate our predictions using data from air quality sensors to ensure accuracy in the upcoming hours. A general workflow of \textsc{FireCOM} can be seen in Figure \ref{figure2}.

By integrating real-time data, GIS tools, and fluid simulations within the Blender platform, \textsc{FireCOM} serves as a reliable digital twin architecture for active fire monitoring and prediction, ultimately enhancing urban safety and public health.

\section{Data}

Real-time fire incident data are procured from the 911 calls' active incident page as updated for over twenty cities. We note that these cities have vastly different schemes for all emergency information, mixing different date formats, missing certain information like coordinates, and calling the same category different things. To deal with this, we created a unified schema with only the needed information. This includes incident name, date, longitude and latitude coordinates, street address, and reporting department. Although latitude and longitude and street address are translatable to one another, we included both to reduce any geocoding or confusion on the part of the end-user, whose interest is in the street address. We, however, only need the exact coordinates to display on a map. From all of our cities, we manually annotated what each live incident feed returned and created scripts to convert all the different formats we received back, including HTML tables, RSS feeds, plain text files, and JSON documents, into our format. An example of such a script is provided in Algorithm~\ref{algo:script}.

\begin{algorithm}
	\caption{Urban Fire Retrieval} 
	\begin{algorithmic}[1]
		\For {$city=1,2,\ldots$}
				\State Fetch city data from permanent live URL
				\State Based on city, convert to custom format (JSON)
			    \State Clean data with timestamps and geolocations
                    \State Store data in date-based structure
		\EndFor
	\end{algorithmic} \label{algo:script}
\end{algorithm}

After variations of this script run for each and every annotated city's live data, we compare the latest data with the same information retrieved on the previous request, actively updating which fires are still burning and which ones have been reportedly fixed. With this information, we track active fires in all reported cities. We additionally determine where most fires in each city are likely to occur by looking at the average amount of fires per region, creating a "fire risk" map of Austin.

\begin{figure}[h]
  \centering
  \includegraphics[width=\linewidth]{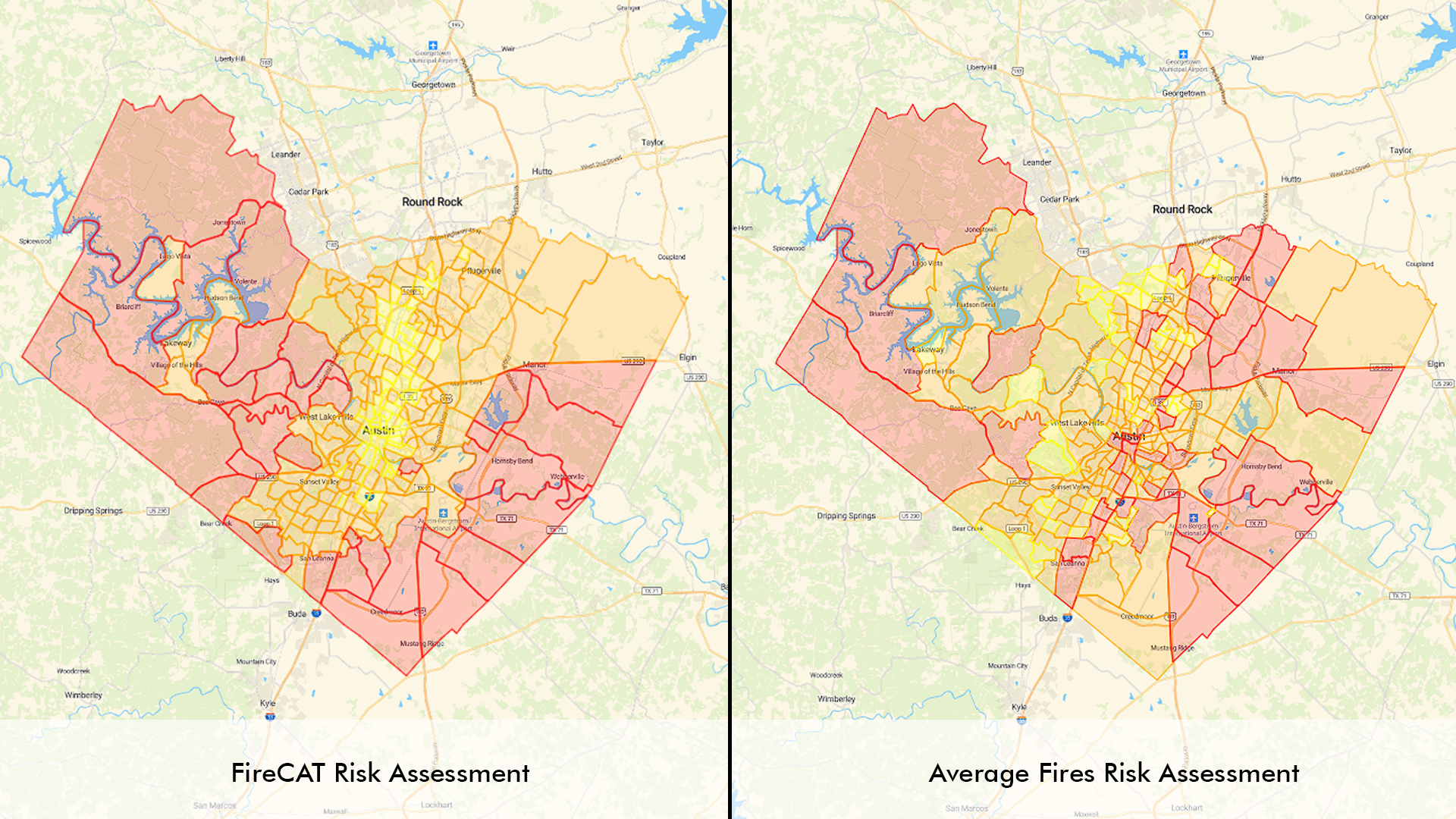}
  \caption{A comparison of FireCAT Wildfire Risk Assessment (left) and our Fire Averages per tract (right).}
  \Description{A comparison of FireCAT Wildfire Risk Assessment and our Fire Averages per tract.}
  \label{fig:fire-risk}
\end{figure}

In 2022, over 20,000 fires occurred in Austin alone. Analyzing this data revealed several trends. By intersecting each fire event with a tract-level map of Austin, we created a "fire risk" analysis map where orange represents normal fire risk, while yellow and red represent below-average and above-average risk, respectively. This map exhibits a correlation with the City of Austin's FireCAT Wildfire Risk Assessment \citep{pohl_2022}, as seen in Figure \ref{fig:fire-risk}, indicating that wildfires remain a significant source of fire incidents. However, it also highlights the elevated probability of fires in downtown Austin. 

\begin{figure}[h]
  \centering
  \includegraphics[width=\linewidth]{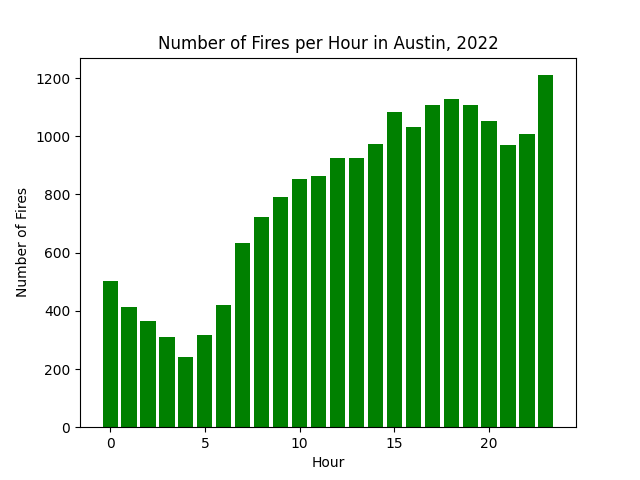}
  \vspace{-4mm}
  \caption{A graph of the number of reported fires by the hour.}
  \Description{A graph of the number of reported fires by the hour.}
  \label{fig:fire-time-trend}
\end{figure}

Our results suggest that most fire incidents occur in the afternoon, typically between 5 p.m. and 10 p.m. (see Figure~\ref{fig:fire-time-trend}). However, the transition between different time periods is not as smooth as one might expect, warranting further investigation into the underlying factors influencing these trends.

Weather service data is incredibly important while surveying smoke impacts, as wind direction and strength, among other weather conditions, ultimately determine the region of impact. As demonstrated by the HRRR-Smoke model for wildfires, weather is perhaps the single most important factor that determines the spatial progress of smoke \citep{HighResolutionSmokeForecasting}. City weather is very well-researched and provided mainly by \url{Weather.gov}, a government service that provides hourly updates of weather conditions across the entire continental United States. We query Weather.gov and selectively retrieve the important information, namely, wind speed, direction, and conditions.

Finally, we create a large data set of time-independent data. This data set includes elevation maps, building footprint data sets, and locations of essential points of interest like hospitals and fire departments. The elevation data is similarly available to the public from OpenTopography, and the rest are publicly available from Open Data Portals for each respective city.

\section{Smoke Prediction and Validation}

\subsection{Methods}

To predict smoke output from various tracked fires, we initially employed a Gaussian smoke prediction model based on the VSmoke particulate matter concentration program developed by the Georgia Forestry Commission (GFC) \citep{Lavdas, Lavdas_1996}. This model calculates smoke dispersion using a normal probability distribution, which is common for ground-level fires \citep{Bosanquet1936TheSO}.

We gathered real-time weather information, such as wind speed, wind direction, and humidity, from Weather.gov and NOAA’s HRRR model to input into the VSmoke model. From the live fire information, we can make average assumptions based on the size of each fire, such as an “appliance fire” versus a “brush fire,” what type of fuel the fire uses, as well as tweak other qualities of the active fire. 

However, it's important to note that VSmoke doesn't account for the surrounding buildings' geometry. To address this, we moved to a 3D city model using Blender and OpenStreetMaps 3D \citep{Over2010GeneratingW3}, which allowed us to implement a more accurate smoke prediction model using MantaFlow \citep{10.1145/3072959.3073643,10.1145/3197517.3201304}.

\begin{figure}[h]
  \centering
  \includegraphics[width=\linewidth]{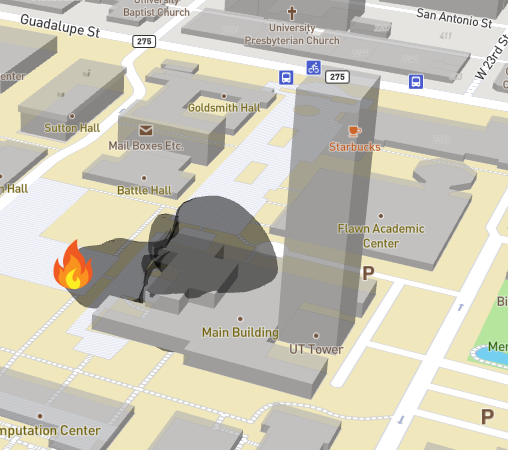}
  \caption{A supervised smoke fluid simulation in-browser.}
  \Description{A supervised smoke fluid simulation in-browser.} \label{fig:summary}
  \label{figure5}
\end{figure}

To do this, Blender was used, alongside OpenStreetMaps 3D, which is a generated mesh of an entire city using building footprints and available height information to create approximate models of each structure \citep{Over2010GeneratingW3}. We also overlay it on an accurate topographic height map of Austin using OpenTopography, down to 30m precision. With the surrounding environment in-place, we created a new BlenderPy script to spawn a smoke fluid simulation at the exact longitude and latitude within the “3D City”, with parameters for weather and fire matching those input into VSmoke. In this way, we run a smoke simulation in a near-realistic digital twin of any city. An example of this visualization, converted from Blender to a browser-view, can be seen in Figure \ref{figure5}.

\subsection{Results}

\begin{figure}[h]
  \centering
  \includegraphics[width=\linewidth]{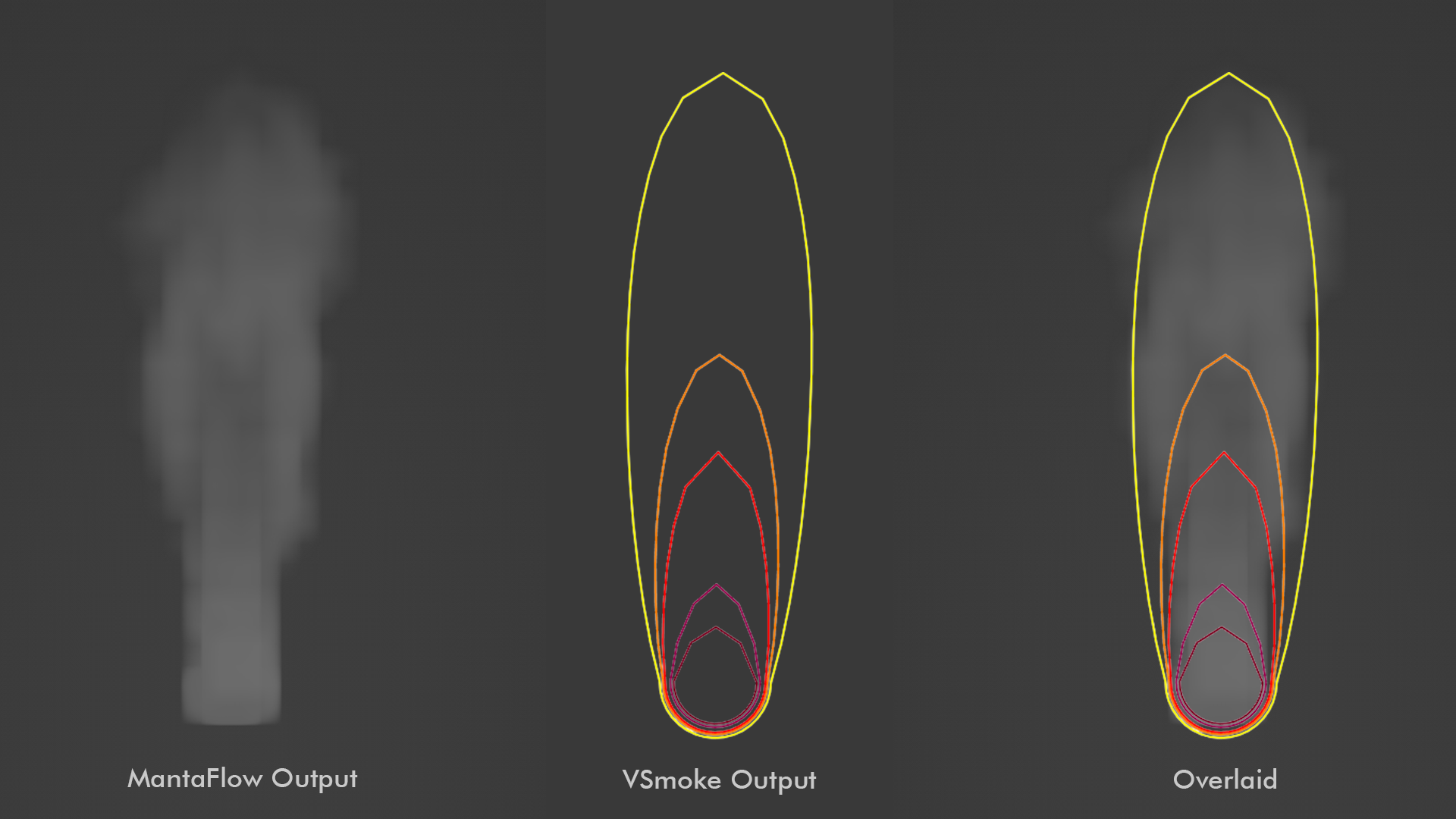}
  \caption{A comparison of the VSmoke and MantaFlow predicted smoke outputs.}
  \Description{A comparison of the VSmoke and MantaFlow predicted smoke outputs.}
  \label{fig:comparison}
\end{figure}

Using the VSmoke model as a reference, we tuned the MantaFlow fluid simulation parameters to match the VSmoke prediction area. Figure~\ref{fig:comparison} shows a comparison between the VSmoke and MantaFlow predicted smoke outputs.

Typically, smoke systems are validated from air sensors, satellite imagery, and meteorology stations \citep{HighResolutionSmokeForecasting}. Since satellite imagery does not work well for fires in an urban context, as well as that meteorology stations were often sparse and located far from cities, we assessed the accuracy of our smoke predictions by comparing them with air sensor data. We collaborated with researchers from the Texas Advanced Computing Center (TACC) to evaluate the overlap between our generated smoke trails and PurpleAir sensor data and found that sensors were too sparse to detect urban smoke impacts. Instead, we used a validation process by case studies with the City of Austin, where we were able to use prescribed burns and set up local air sensors ahead of time to test the accuracy of our model.

In terms of pipeline performance, our system proved to be efficient in generating smoke predictions and displaying them in the browser. Upon receiving a notification from the active fire query API, the system takes less than a second to generate an approximate VSmoke prediction for up to three hours in advance (not considering city geometry) and communicate that result in-browser. For the more accurate MantaFlow 3D smoke predictions, which take into account the city geometry, the pipeline takes approximately thirty seconds to generate predictions for up to three hours in advance. This timely generation and display of smoke predictions can be critical in facilitating rapid decision-making and response efforts in urban fire scenarios.

\subsection{Discussion}

Our study resulted in two models: the VSmoke-based model, which is computationally efficient but less accurate, and the MantaFlow-based model, which is computationally intensive but provides more accurate results by considering complex city geometry. 

As for validation, our partners at TACC demonstrated that air sensors within our cities were too sparse, never being even within a mile of a documented fire. On the few collisions, they suggested that our findings were accurate, as air sensors reported lower than average values within the predicted smoke area. However, it's important to note that air quality sensors in Austin and major cities are too sparse to detect acute events like fires, even a decade after a similar issue was reported by NASA \citep{nasa}. NASA's solutions to this problem of "sparse data" are generally smoothing of data points over large regions, which is still not suitable for the smoke from urban fires, which would be a spike in a small area.

Instead, we used experimental data provided from a series of case studies conducted with the City of Austin. The Austin Fire Department, like many fire departments, prescribes fires weeks in advance that need to occur for firefighter training, brush removal, or forest clearing. Knowing the location of a future fire, we could set up our air sensors downwind of the location and detect if our model was accurate. We had predicted that air sensors that did not change under downwind conditions were often behind buildings and other obstructive geometry, which could be accounted for with our 3D smoke model. The MantaFlow-based model demonstrated promising accuracy in our case study with the City of Austin, finding that the air quality of downwind regions correlates highly with our predicted smoke impact.

Despite the positive results, our study faced limitations due to the sparsity of air quality sensors in Austin and other major cities. We recommend future research in this area to consider deploying low-cost air sensors across cities for improved detection and reporting of immediate threats to air quality.

However, we also recognize that some of our tracked cities, like Los Angeles, had thousands of air sensors, yet exhibited the same behaviour. For this reason, we also recommend the development of novel air quality tracking techniques, like geostationary satellite detection, currently being worked on by NASA ARSET \citep{ARSET2022}. With tools like these, validation of fire and smoke digital twins like \textsc{FireCOM} will become much easier.

\section{Key challenges, solutions, and innovations} \label{sec:key}

A key challenge in conducting this research was the relative scarcity of smoke prediction models. While models do, to an extent, exist, they are often very specific or meant for non-scientific usage. For example, the EPA’s smoke model predicts long continuous regions of smoke across North America, without really offering much useful information in terms of actual air quality impacts on communities. Another example is the fluid simulations of game engines, such Unreal Engine, which are meant to be optimized for visualization rather than any consideration to accurately portray how smoke and fire act in life. For these reasons, we chose to use smoke models from VSmoke from the Georgia Forestry Commission and MantaFlow provided by the Technical University of Munich. 

Another challenge to overcome was the inability to present these models in any direct way to viewers. Both fluid simulations and real-time active maps are rather costly to compute and maintain on a running basis, and doing these smoke predictions on-demand to display immediately within a live map required some technical innovations. Displaying fluid simulations interactively to users is a pipe dream, as well, as they take a powerful computer to compute, yet alone visualize. We needed to make the technology accessible at the whim of a smartphone’s web browser. To this end, we created APIs for both our 2D and 3D smoke pipelines, as well as used some tricks to display the output of each respective simulation. 

Firstly, we rewrote a lot of GFC’s codebase to optimize it for newer computers, as well as run parallel smoke simulations at the request of an API. The program outputs Keyhole Markup (.KML) files to visualize each 2D smoke simulation, which we spaced out at one hour, two hour, and three hour predicted simulations. Secondly, we created an API to run fluid simulations for fires on-demand in Blender. While the model expanded for a full three hours in Blender, we timestamped and exported the one, two, and three hour marks to Filmbox (.FBX) object files, which can be rendered in-browser. 

\begin{figure}[h]
  \centering
  \includegraphics[width=\linewidth]{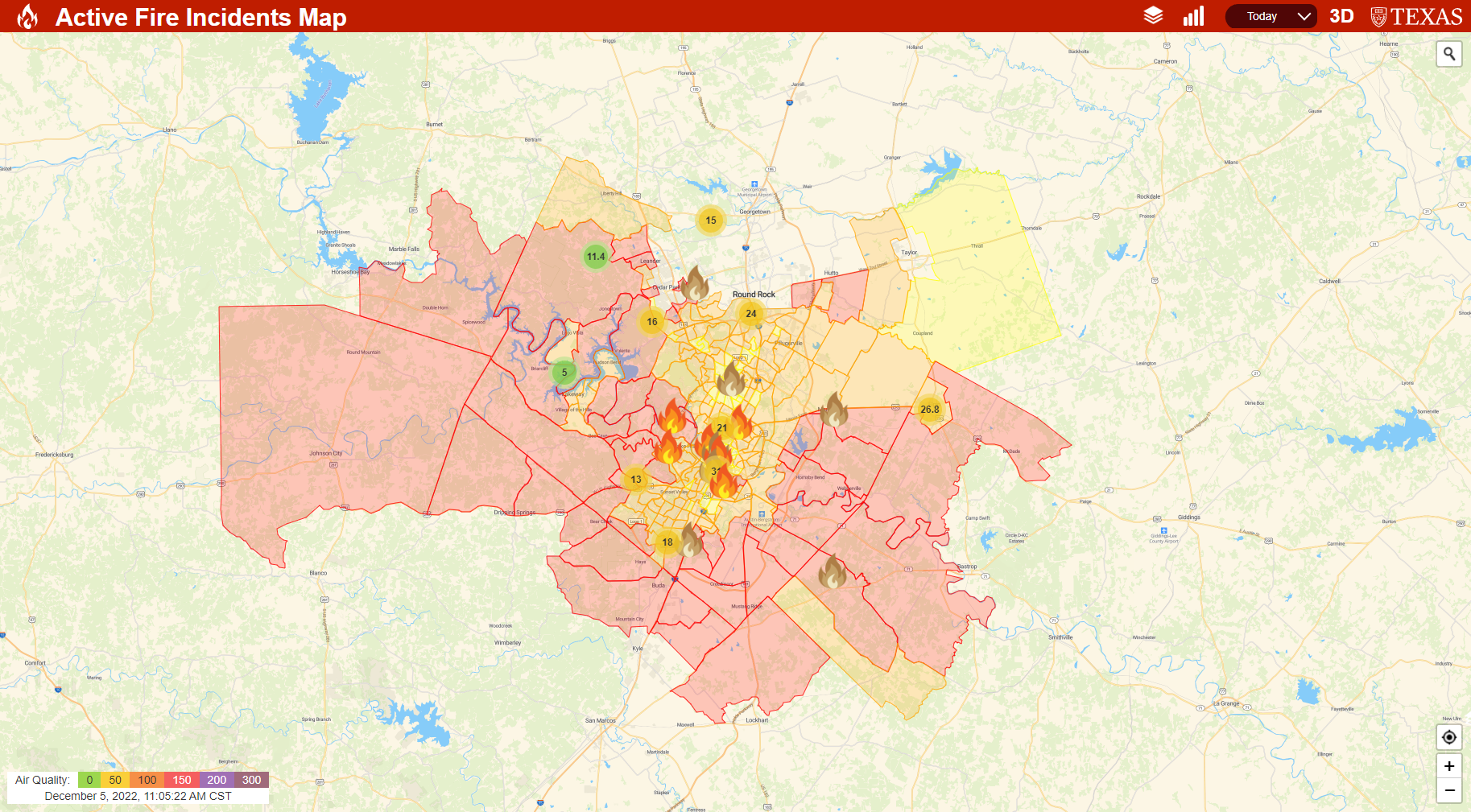}
  \caption{Our 2D real-time fire and smoke map in-browser.}
  \Description{Our 2D real-time fire and smoke map in-browser.}
  \label{figure7}
\end{figure}

\begin{figure}[h]
  \centering
  \includegraphics[width=\linewidth]{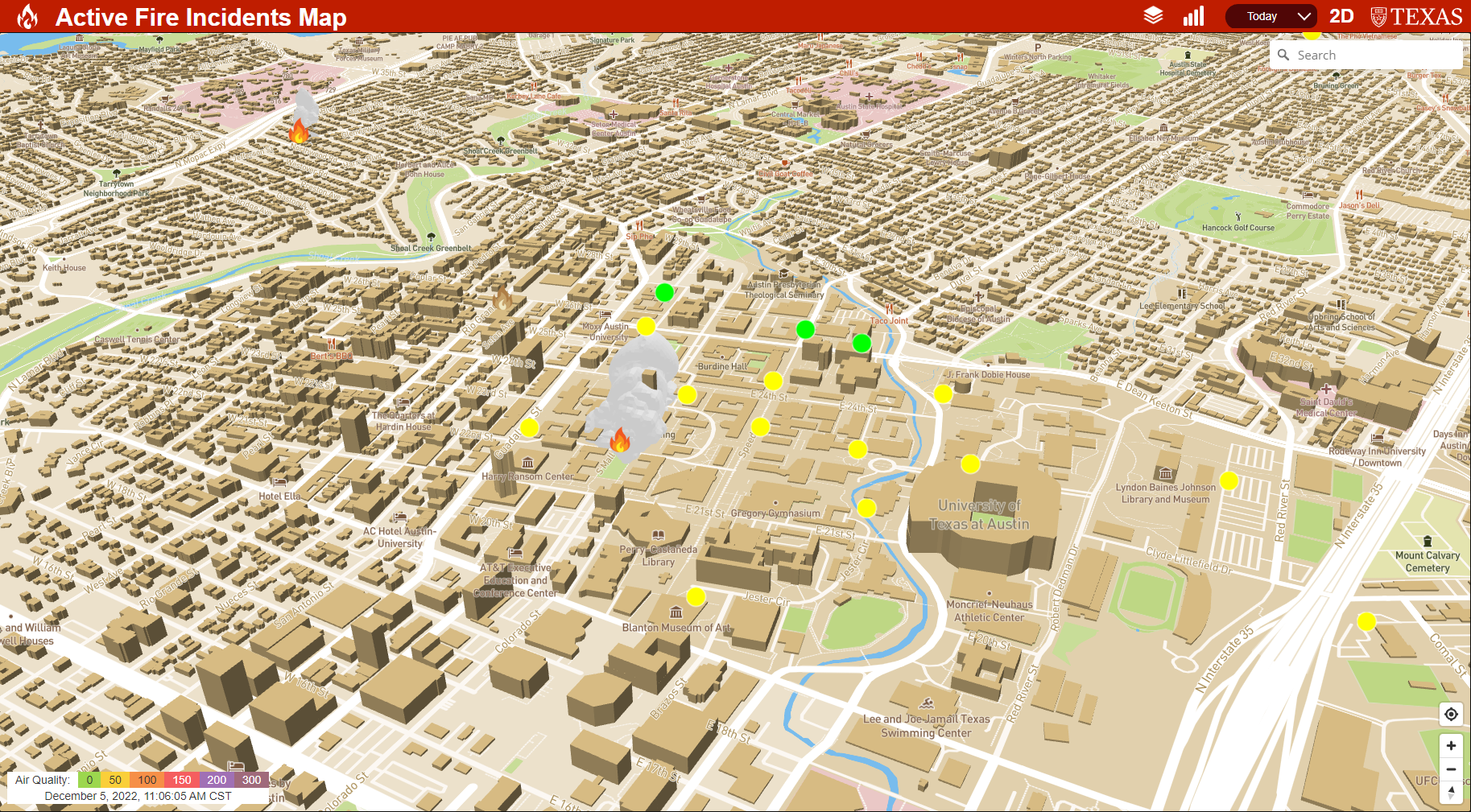}
  \caption{Our 3D real-time fire and smoke map in-browser.}
  \Description{Our 3D real-time fire and smoke map in-browser.}
  \label{figure8}
\end{figure}

With the smoke generated in six different iterations, we just needed to display our results on each browser. To reduce confusion between the two separate models, we displayed VSmoke’s simulations on a standard map labelled “2D”, while presenting the MantaFlow simulations on a vector-tiled 3D map, courtesy of MapBox GL, which quickly renders the same OpenStreetMap 3D tiles from Blender in-browser. From this, we just used some extension libraries, like Threebox.js, to display our smoke simulations within the 3D space from the fire’s origin, all without requiring anything other than a machine capable of viewing a model, which each computer and smartphone is capable of. Our computationally efficient VSmoke-enabled 2D map can be seen in Figure \ref{figure7}, while our fluid-simulated MantaFlow 3D map can be seen in Figure \ref{figure8}. We also overlay a dynamically updating "fire risk" layer, which changes as the average amount of fires per tract in Austin updates day by day.

The novelty of our work lends itself to three areas: (1) data integration, (2) shifting from 2D to 3D, and (3) use of public and generalized data. We integrated information from a multitude of resources in creating this real-time model, including building footprints, air sensors, wind and weather updates, as well as census data and points of interest around the city. With all of this data, we can easily begin to generate smoke models across the city, but with only an approximate accuracy, given the complexity of an urban environment. Shifting to a 3D perspective of a city, complete with topography and building geometry, allows for a more realistic and better detailed understanding of how each fire falloff will develop, and where exactly it will affect, allowing both citizens and firefighters to make informed decisions when dealing with fire in their communities. Finally, our work was done completely with public data, and alongside our released open-source code, makes this model completely generalizable and replicable to any city. 

\section{Conclusions}

On-demand smoke simulations for urban fires were never attempted because of the complex geometry and insights needed to perform such a task, but with our model, one can easily predict both approximate smoke outcomes, as well as hyper-realistic smoke fluid simulations for any fire. With our APIs, researchers can begin to predict future fires, as well as analyze past fires for any city. With our map, citizens can become well-informed of the air quality risks posed by urban fires around their city, as well as better understand how smoke might impact their nearby communities. We could potentially serve as an early warning system for lowered air quality and smoke in areas of high risk, where exposure to reduced PM2.5 could endanger lives and hinder child development.

In the near future, we expect to see our model used to conduct air quality research into respiratory diseases. We can detect if there is a correlation between areas of high urban fire risk alongside areas of high COPD, Asthma, and other respiratory issues. On top of this, we wish to see our model used in growing digital twins of Smart City environments, whereas every aspect of a city can be visualized, predicted, and optimized to save lives and serve communities better. 

Within a full digital twin model, our current data would be a marginal list of the information necessary. A full twin might take advantage of the traffic conditions on a right or left lane to indicate which side of the road a burning car might be on, or even make estimations based on the total aggregation of phone signals in an area about how many people might be potentially affected by a smoke falloff. For future research in this area, we would encourage the use of \textit{Big Data} in retrieving all relevant local conditions and creating a more full-scale model.

\section*{Funding}
This research is supported by the Bridging Barriers Initiative Good Systems Grand Challenge at The University of Texas at Austin, the City of Austin (UTA19-000382), National Science Foundation (2043060, 2133302, 1952193, 2125858, 2236305), NSF-GOLD (2228205), and CSE-OCE (1835739).

\begin{acks}
The authors extend their sincere gratitude to Dan Chen (Georgia Forestry Association), Marc Coudert (Office of Resilience, City of Austin), Braniff Davis (Austin Fire Department), and Chief Joel Baker (Austin Fire Department) for supporting our project. The authors also acknowledge an interagency agreement between UT and the City of Austin through the Bridging Barriers Initiative.
\end{acks}

\bibliographystyle{ACM-Reference-Format}
\bibliography{main}


\begin{thebibliography}{38}


\ifx \showCODEN    \undefined \def \showCODEN     #1{\unskip}     \fi
\ifx \showDOI      \undefined \def \showDOI       #1{#1}\fi
\ifx \showISBNx    \undefined \def \showISBNx     #1{\unskip}     \fi
\ifx \showISBNxiii \undefined \def \showISBNxiii  #1{\unskip}     \fi
\ifx \showISSN     \undefined \def \showISSN      #1{\unskip}     \fi
\ifx \showLCCN     \undefined \def \showLCCN      #1{\unskip}     \fi
\ifx \shownote     \undefined \def \shownote      #1{#1}          \fi
\ifx \showarticletitle \undefined \def \showarticletitle #1{#1}   \fi
\ifx \showURL      \undefined \def \showURL       {\relax}        \fi
\providecommand\bibfield[2]{#2}
\providecommand\bibinfo[2]{#2}
\providecommand\natexlab[1]{#1}
\providecommand\showeprint[2][]{arXiv:#2}

\bibitem[Alarie(2002)]%
        {alarie_2002}
\bibfield{author}{\bibinfo{person}{Yves Alarie}.}
  \bibinfo{year}{2002}\natexlab{}.
\newblock \showarticletitle{Toxicity of fire smoke}.
\newblock \bibinfo{journal}{\emph{Critical Reviews in Toxicology}}
  \bibinfo{volume}{32}, \bibinfo{number}{4} (\bibinfo{year}{2002}),
  \bibinfo{pages}{259–289}.
\newblock
\urldef\tempurl%
\url{https://doi.org/10.1080/20024091064246}
\showDOI{\tempurl}


\bibitem[Allen and Melgar(2019)]%
        {predicteq}
\bibfield{author}{\bibinfo{person}{Richard Allen} {and} \bibinfo{person}{Diego
  Melgar}.} \bibinfo{year}{2019}\natexlab{}.
\newblock \showarticletitle{Earthquake Early Warning: Advances, Scientific
  Challenges, and Societal Needs}.
\newblock \bibinfo{journal}{\emph{Annual Review of Earth and Planetary
  Sciences}}  \bibinfo{volume}{47} (\bibinfo{date}{05} \bibinfo{year}{2019}).
\newblock
\urldef\tempurl%
\url{https://doi.org/10.1146/annurev-earth-053018-060457}
\showDOI{\tempurl}


\bibitem[Antwi-Agyakwa et~al\mbox{.}(2023)]%
        {predictfloods}
\bibfield{author}{\bibinfo{person}{Kwasi Antwi-Agyakwa},
  \bibinfo{person}{Mawuli Afenyo}, {and} \bibinfo{person}{Donatus Angnuureng}.}
  \bibinfo{year}{2023}\natexlab{}.
\newblock \showarticletitle{Know to Predict, Forecast to Warn: A Review of
  Flood Risk Prediction Tools}.
\newblock \bibinfo{journal}{\emph{Water}}  \bibinfo{volume}{15}
  (\bibinfo{date}{01} \bibinfo{year}{2023}).
\newblock
\urldef\tempurl%
\url{https://doi.org/10.3390/w15030427}
\showDOI{\tempurl}


\bibitem[Bosanquet and Pearson(1936)]%
        {Bosanquet1936TheSO}
\bibfield{author}{\bibinfo{person}{C.~H. Bosanquet} {and}
  \bibinfo{person}{James~L. Pearson}.} \bibinfo{year}{1936}\natexlab{}.
\newblock \showarticletitle{The spread of smoke and gases from chimneys}.
\newblock \bibinfo{journal}{\emph{Transactions of The Faraday Society}}
  \bibinfo{volume}{32} (\bibinfo{year}{1936}), \bibinfo{pages}{1249--1263}.
\newblock


\bibitem[Bush et~al\mbox{.}(2000)]%
        {firepesticides}
\bibfield{author}{\bibinfo{person}{Parshall~B. Bush},
  \bibinfo{person}{Daniel~G. Neary}, {and} \bibinfo{person}{Charles~K.
  McMahon}.} \bibinfo{year}{2000}\natexlab{}.
\newblock \bibinfo{title}{Fire and Pesticides: A Review of Air Quality
  Considerations}.
\newblock
\newblock


\bibitem[Carlin et~al\mbox{.}(2015)]%
        {Asbestos}
\bibfield{author}{\bibinfo{person}{Danielle Carlin}, \bibinfo{person}{Theodore
  Larson}, \bibinfo{person}{Jean Pfau}, \bibinfo{person}{Stephen Gavett},
  \bibinfo{person}{Arti Shukla}, \bibinfo{person}{Aubrey Miller}, {and}
  \bibinfo{person}{Ronald Hines}.} \bibinfo{year}{2015}\natexlab{}.
\newblock \showarticletitle{Current Research and Opportunities to Address
  Environmental Asbestos Exposures}.
\newblock \bibinfo{journal}{\emph{Environmental health perspectives}}
  \bibinfo{volume}{123} (\bibinfo{date}{08} \bibinfo{year}{2015}),
  \bibinfo{pages}{A194--A197}.
\newblock
\urldef\tempurl%
\url{https://doi.org/10.1289/ehp.1409662}
\showDOI{\tempurl}


\bibitem[Chow et~al\mbox{.}(2022)]%
        {HighResolutionSmokeForecasting}
\bibfield{author}{\bibinfo{person}{Fotini~Katopodes Chow},
  \bibinfo{person}{Katelyn~A. Yu}, \bibinfo{person}{Alexander Young},
  \bibinfo{person}{Eric James}, \bibinfo{person}{Georg~A. Grell},
  \bibinfo{person}{Ivan Csiszar}, \bibinfo{person}{Marina Tsidulko},
  \bibinfo{person}{Saulo Freitas}, \bibinfo{person}{Gabriel Pereira},
  \bibinfo{person}{Louis Giglio}, \bibinfo{person}{Mariel~D. Friberg}, {and}
  \bibinfo{person}{Ravan Ahmadov}.} \bibinfo{year}{2022}\natexlab{}.
\newblock \showarticletitle{High-Resolution Smoke Forecasting for the 2018 Camp
  Fire in California}.
\newblock \bibinfo{journal}{\emph{Bulletin of the American Meteorological
  Society}} \bibinfo{volume}{103}, \bibinfo{number}{6} (\bibinfo{year}{2022}),
  \bibinfo{pages}{E1531 -- E1552}.
\newblock
\urldef\tempurl%
\url{https://doi.org/10.1175/BAMS-D-20-0329.1}
\showDOI{\tempurl}


\bibitem[Chu and Thuerey(2017)]%
        {10.1145/3072959.3073643}
\bibfield{author}{\bibinfo{person}{Mengyu Chu} {and} \bibinfo{person}{Nils
  Thuerey}.} \bibinfo{year}{2017}\natexlab{}.
\newblock \showarticletitle{Data-Driven Synthesis of Smoke Flows with CNN-Based
  Feature Descriptors}.
\newblock \bibinfo{journal}{\emph{ACM Trans. Graph.}} \bibinfo{volume}{36},
  \bibinfo{number}{4}, Article \bibinfo{articleno}{69} (\bibinfo{date}{jul}
  \bibinfo{year}{2017}), \bibinfo{numpages}{14}~pages.
\newblock
\showISSN{0730-0301}
\urldef\tempurl%
\url{https://doi.org/10.1145/3072959.3073643}
\showDOI{\tempurl}


\bibitem[Ferreira et~al\mbox{.}(2014)]%
        {urbanfirerisk}
\bibfield{author}{\bibinfo{person}{Tiago Ferreira}, \bibinfo{person}{Romeu
  Vicente}, \bibinfo{person}{Raimundo Mendes~Silva}, \bibinfo{person}{H.
  Varum}, \bibinfo{person}{A. Costa}, {and} \bibinfo{person}{Rui Maio}.}
  \bibinfo{year}{2014}\natexlab{}.
\newblock \showarticletitle{Urban Fire Risk: evaluation and emergency
  planning}.
\newblock
\urldef\tempurl%
\url{https://doi.org/10.13140/2.1.5053.6648}
\showDOI{\tempurl}


\bibitem[Gil et~al\mbox{.}(2021)]%
        {10.1145/3453172}
\bibfield{author}{\bibinfo{person}{Yolanda Gil}, \bibinfo{person}{Daniel
  Garijo}, \bibinfo{person}{Deborah Khider}, \bibinfo{person}{Craig~A.
  Knoblock}, \bibinfo{person}{Varun Ratnakar}, \bibinfo{person}{Maximiliano
  Osorio}, \bibinfo{person}{Hern\'{a}n Vargas}, \bibinfo{person}{Minh Pham},
  \bibinfo{person}{Jay Pujara}, \bibinfo{person}{Basel Shbita},
  \bibinfo{person}{Binh Vu}, \bibinfo{person}{Yao-Yi Chiang},
  \bibinfo{person}{Dan Feldman}, \bibinfo{person}{Yijun Lin},
  \bibinfo{person}{Hayley Song}, \bibinfo{person}{Vipin Kumar},
  \bibinfo{person}{Ankush Khandelwal}, \bibinfo{person}{Michael Steinbach},
  \bibinfo{person}{Kshitij Tayal}, \bibinfo{person}{Shaoming Xu},
  \bibinfo{person}{Suzanne~A. Pierce}, \bibinfo{person}{Lissa Pearson},
  \bibinfo{person}{Daniel Hardesty-Lewis}, \bibinfo{person}{Ewa Deelman},
  \bibinfo{person}{Rafael Ferreira~Da Silva}, \bibinfo{person}{Rajiv Mayani},
  \bibinfo{person}{Armen~R. Kemanian}, \bibinfo{person}{Yuning Shi},
  \bibinfo{person}{Lorne Leonard}, \bibinfo{person}{Scott Peckham},
  \bibinfo{person}{Maria Stoica}, \bibinfo{person}{Kelly Cobourn},
  \bibinfo{person}{Zeya Zhang}, \bibinfo{person}{Christopher Duffy}, {and}
  \bibinfo{person}{Lele Shu}.} \bibinfo{year}{2021}\natexlab{}.
\newblock \showarticletitle{Artificial Intelligence for Modeling Complex
  Systems: Taming the Complexity of Expert Models to Improve Decision Making}.
\newblock \bibinfo{journal}{\emph{ACM Trans. Interact. Intell. Syst.}}
  \bibinfo{volume}{11}, \bibinfo{number}{2}, Article \bibinfo{articleno}{11}
  (\bibinfo{date}{jul} \bibinfo{year}{2021}), \bibinfo{numpages}{49}~pages.
\newblock
\showISSN{2160-6455}
\urldef\tempurl%
\url{https://doi.org/10.1145/3453172}
\showDOI{\tempurl}


\bibitem[Goswami et~al\mbox{.}(2020)]%
        {airpol}
\bibfield{author}{\bibinfo{person}{Meera Goswami}, \bibinfo{person}{Dalip
  Mansotra}, \bibinfo{person}{Shivalika Sharma}, \bibinfo{person}{Gaurav Pant},
  {and} \bibinfo{person}{Prakash Joshi}.} \bibinfo{year}{2020}\natexlab{}.
\newblock \bibinfo{booktitle}{\emph{EFFECTS OF AIR POLLUTION ON HUMAN HEALTH}}.
\newblock \bibinfo{pages}{135--144}.
\newblock
\showISBNx{9788170196587}


\bibitem[Himoto and Tanaka(2003)]%
        {urbanfiremodel}
\bibfield{author}{\bibinfo{person}{Keisuke Himoto} {and}
  \bibinfo{person}{Takeyoshi Tanaka}.} \bibinfo{year}{2003}\natexlab{}.
\newblock \showarticletitle{A Physically-Based Model for Urban Fire Spread}.
\newblock \bibinfo{journal}{\emph{Fire Safety Science}}  \bibinfo{volume}{7}
  (\bibinfo{date}{01} \bibinfo{year}{2003}), \bibinfo{pages}{129--140}.
\newblock
\urldef\tempurl%
\url{https://doi.org/10.3801/IAFSS.FSS.7-129}
\showDOI{\tempurl}


\bibitem[{Land Atmosphere Near Real-Time Capability For EOS Fire Information
  For Resource Management System}(2021)]%
        {https://doi.org/10.5067/firms/modis/mcd14dl.nrt.0061}
\bibfield{author}{\bibinfo{person}{{Land Atmosphere Near Real-Time Capability
  For EOS Fire Information For Resource Management System}}.}
  \bibinfo{year}{2021}\natexlab{}.
\newblock \bibinfo{title}{MODIS/Aqua+Terra Thermal Anomalies/Fire locations 1km
  FIRMS V006 NRT (Vector data)}.
\newblock
\newblock
\urldef\tempurl%
\url{https://doi.org/10.5067/FIRMS/MODIS/MCD14DL.NRT.0061}
\showDOI{\tempurl}


\bibitem[Larkin et~al\mbox{.}(2009)]%
        {bluesky}
\bibfield{author}{\bibinfo{person}{Narasimhan Larkin}, \bibinfo{person}{Susan
  O'Neill}, \bibinfo{person}{Robert Solomon}, \bibinfo{person}{Sean Raffuse},
  \bibinfo{person}{Tara Strand}, \bibinfo{person}{Dana Coe},
  \bibinfo{person}{Candace Krull}, \bibinfo{person}{Miriam Rorig},
  \bibinfo{person}{Janice Peterson}, {and} \bibinfo{person}{Sue Ferguson}.}
  \bibinfo{year}{2009}\natexlab{}.
\newblock \showarticletitle{The BlueSky smoke modeling framework}.
\newblock \bibinfo{journal}{\emph{Int. J. Wildland Fire}}  \bibinfo{volume}{18}
  (\bibinfo{date}{01} \bibinfo{year}{2009}).
\newblock
\urldef\tempurl%
\url{https://doi.org/10.1071/WF07086}
\showDOI{\tempurl}


\bibitem[Lavdas(1996a)]%
        {Lavdas}
\bibfield{author}{\bibinfo{person}{L.~G. Lavdas}.}
  \bibinfo{year}{1996}\natexlab{a}.
\newblock \showarticletitle{{Improving Control of Smoke from Prescribed Fire
  Using Low Visibility Occurrence Risk Index}}.
\newblock \bibinfo{journal}{\emph{Southern Journal of Applied Forestry}}
  \bibinfo{volume}{20}, \bibinfo{number}{1} (\bibinfo{date}{02}
  \bibinfo{year}{1996}), \bibinfo{pages}{10--14}.
\newblock
\showISSN{0148-4419}
\urldef\tempurl%
\url{https://doi.org/10.1093/sjaf/20.1.10}
\showDOI{\tempurl}
\showeprint{https://academic.oup.com/sjaf/article-pdf/20/1/10/23457356/sjaf0010.pdf}


\bibitem[Lavdas(1996b)]%
        {Lavdas_1996}
\bibfield{author}{\bibinfo{person}{Leonidas~G. Lavdas}.}
  \bibinfo{year}{1996}\natexlab{b}.
\newblock \bibinfo{booktitle}{\emph{Program {VSMOKE}--Users Manual}}.
\newblock \bibinfo{type}{{T}echnical {R}eport}.
\newblock
\urldef\tempurl%
\url{https://doi.org/10.2737/srs-gtr-6}
\showDOI{\tempurl}


\bibitem[Li et~al\mbox{.}(2012)]%
        {LI2012690}
\bibfield{author}{\bibinfo{person}{Junmei Li}, \bibinfo{person}{Shanshan Liu},
  \bibinfo{person}{Yanfeng Li}, \bibinfo{person}{Chao Chen},
  \bibinfo{person}{Xuan Liu}, {and} \bibinfo{person}{Chenchen Yin}.}
  \bibinfo{year}{2012}\natexlab{}.
\newblock \showarticletitle{Experimental Study of Smoke Spread in Titled Urban
  Traffic Tunnels Fires}.
\newblock \bibinfo{journal}{\emph{Procedia Engineering}}  \bibinfo{volume}{45}
  (\bibinfo{year}{2012}), \bibinfo{pages}{690--694}.
\newblock
\showISSN{1877-7058}
\urldef\tempurl%
\url{https://doi.org/10.1016/j.proeng.2012.08.224}
\showDOI{\tempurl}
\newblock
\shownote{2012 International Symposium on Safety Science and Technology}.


\bibitem[Li and Davidson(2013)]%
        {Li2013}
\bibfield{author}{\bibinfo{person}{Sizheng Li} {and} \bibinfo{person}{Rachel~A.
  Davidson}.} \bibinfo{year}{2013}\natexlab{}.
\newblock \showarticletitle{Parametric study of urban fire spread using an
  urban fire simulation model with fire department suppression}.
\newblock \bibinfo{journal}{\emph{Fire Safety Journal}}  \bibinfo{volume}{61}
  (\bibinfo{date}{Oct.} \bibinfo{year}{2013}), \bibinfo{pages}{217--225}.
\newblock
\urldef\tempurl%
\url{https://doi.org/10.1016/j.firesaf.2013.09.017}
\showDOI{\tempurl}


\bibitem[Mandel et~al\mbox{.}(2014)]%
        {sfire}
\bibfield{author}{\bibinfo{person}{J. Mandel}, \bibinfo{person}{Shai Amram},
  \bibinfo{person}{J. Beezley}, \bibinfo{person}{Guy Kelman},
  \bibinfo{person}{Adam Kochanski}, \bibinfo{person}{Volodymyr Kondratenko},
  \bibinfo{person}{B. Lynn}, \bibinfo{person}{B. Regev}, {and}
  \bibinfo{person}{Martin Vejmelka}.} \bibinfo{year}{2014}\natexlab{}.
\newblock \showarticletitle{Recent advances and applications of WRF–SFIRE}.
\newblock \bibinfo{journal}{\emph{Natural Hazards and Earth System Science}}
  \bibinfo{volume}{14} (\bibinfo{date}{10} \bibinfo{year}{2014}),
  \bibinfo{pages}{2829--2845}.
\newblock
\urldef\tempurl%
\url{https://doi.org/10.5194/nhess-14-2829-2014}
\showDOI{\tempurl}


\bibitem[Matson and Dozier(1981)]%
        {Dozier}
\bibfield{author}{\bibinfo{person}{M. Matson} {and} \bibinfo{person}{Jeff
  Dozier}.} \bibinfo{year}{1981}\natexlab{}.
\newblock \showarticletitle{Identification of subresolution high temperature
  sources using thermal IR sensor}.
\newblock \bibinfo{journal}{\emph{Photogrammetric Engineering and Remote
  Sensing}}  \bibinfo{volume}{47} (\bibinfo{date}{01} \bibinfo{year}{1981}),
  \bibinfo{pages}{1311--1318}.
\newblock


\bibitem[Nanjappan et~al\mbox{.}(2021)]%
        {wildfirehealth}
\bibfield{author}{\bibinfo{person}{Eshwari Nanjappan}, \bibinfo{person}{Emily
  Sullo}, \bibinfo{person}{Srijain Shrestha}, \bibinfo{person}{Shara Thomas},
  {and} \bibinfo{person}{Elysée Nouvet}.} \bibinfo{year}{2021}\natexlab{}.
\newblock \showarticletitle{Californian Wildfires and Associated Human Health
  Outcomes: An Epidemiological Scoping Review of the Creative Commons
  Attribution License (CC BY 4.0)}.
\newblock   \bibinfo{volume}{5} (\bibinfo{date}{07} \bibinfo{year}{2021}),
  \bibinfo{pages}{944--953}.
\newblock


\bibitem[{NASA Applied Remote Sensing Training Program (ARSET)}(2022)]%
        {ARSET2022}
\bibfield{author}{\bibinfo{person}{{NASA Applied Remote Sensing Training
  Program (ARSET)}}.} \bibinfo{year}{2022}\natexlab{}.
\newblock \bibinfo{title}{ARSET - Accessing and Analyzing Air Quality Data from
  Geostationary Satellites}.
\newblock
\newblock
\urldef\tempurl%
\url{http://appliedsciences.nasa.gov/join-mission/training/english/arset-accessing-and-analyzing-air-quality-data-geostationary}
\showURL{%
\tempurl}


\bibitem[Ong et~al\mbox{.}(2022)]%
        {ong2022openems}
\bibfield{author}{\bibinfo{person}{Joshua Ong}, \bibinfo{person}{David
  Kulpanowski}, \bibinfo{person}{Yangxinyu Xie}, \bibinfo{person}{Evdokia
  Nikolova}, {and} \bibinfo{person}{Ngoc~Mai Tran}.}
  \bibinfo{year}{2022}\natexlab{}.
\newblock \showarticletitle{OpenEMS: an open-source Package for Two-Stage
  Stochastic and Robust Optimization for Ambulance Location and Routing with
  Applications to Austin-Travis County EMS Data}.
\newblock \bibinfo{journal}{\emph{arXiv preprint arXiv:2201.11208}}
  (\bibinfo{year}{2022}).
\newblock


\bibitem[Over et~al\mbox{.}(2010)]%
        {Over2010GeneratingW3}
\bibfield{author}{\bibinfo{person}{Martin Over}, \bibinfo{person}{Arne
  Schilling}, \bibinfo{person}{S. Neubauer}, {and} \bibinfo{person}{Alexander
  Zipf}.} \bibinfo{year}{2010}\natexlab{}.
\newblock \showarticletitle{Generating web-based 3D City Models from
  OpenStreetMap: The current situation in Germany}.
\newblock \bibinfo{journal}{\emph{Comput. Environ. Urban Syst.}}
  \bibinfo{volume}{34} (\bibinfo{year}{2010}), \bibinfo{pages}{496--507}.
\newblock


\bibitem[Pasch et~al\mbox{.}(2011)]%
        {nasa}
\bibfield{author}{\bibinfo{person}{A. Pasch}, \bibinfo{person}{Patrick Zahn},
  \bibinfo{person}{J. Dewinter}, \bibinfo{person}{M. Haderman},
  \bibinfo{person}{J. White}, \bibinfo{person}{P. Dickerson},
  \bibinfo{person}{Timothy Dye}, {and} \bibinfo{person}{Randall Martin}.}
  \bibinfo{year}{2011}\natexlab{}.
\newblock \showarticletitle{Improve EPA's AIRNow Air Quality Index Maps with
  NASA/NOAA Satellite Data}.
\newblock \bibinfo{journal}{\emph{AGU Fall Meeting Abstracts}}
  (\bibinfo{date}{12} \bibinfo{year}{2011}), \bibinfo{pages}{0376--}.
\newblock


\bibitem[Pohl(2022)]%
        {pohl_2022}
\bibfield{author}{\bibinfo{person}{Kelly Pohl}.}
  \bibinfo{year}{2022}\natexlab{}.
\newblock \bibinfo{title}{Austin wildfire and vulnerable populations tool}.
\newblock
\newblock
\urldef\tempurl%
\url{https://headwaterseconomics.org/wildfire/homes-risk/austin-wildfire-population-risk/}
\showURL{%
\tempurl}


\bibitem[Prichard et~al\mbox{.}(2019)]%
        {wildfireplan}
\bibfield{author}{\bibinfo{person}{Susan Prichard}, \bibinfo{person}{Sim~N.
  Larkin}, \bibinfo{person}{Roger Ottmar}, \bibinfo{person}{Nancy~H.F. French},
  \bibinfo{person}{Kirk Baker}, \bibinfo{person}{Tim Brown},
  \bibinfo{person}{Craig Clements}, \bibinfo{person}{Matt Dickinson},
  \bibinfo{person}{Andrew Hudak}, \bibinfo{person}{Adam Kochanski},
  \bibinfo{person}{Rod Linn}, \bibinfo{person}{Yongqiang Liu},
  \bibinfo{person}{Brian Potter}, \bibinfo{person}{William Mell},
  \bibinfo{person}{Danielle Tanzer}, \bibinfo{person}{Shawn Urbanski}, {and}
  \bibinfo{person}{Adam Watts}.} \bibinfo{year}{2019}\natexlab{}.
\newblock \showarticletitle{The fire and smoke model evaluation experiment—a
  plan for integrated, large fire–atmosphere field campaigns}.
\newblock \bibinfo{journal}{\emph{Atmosphere. 10(2): 66-.}}
  \bibinfo{volume}{10}, \bibinfo{number}{2} (\bibinfo{year}{2019}).
\newblock
\urldef\tempurl%
\url{https://doi.org/10.3390/atmos10020066}
\showDOI{\tempurl}


\bibitem[Roy et~al\mbox{.}(2006)]%
        {Roy2006TheUO}
\bibfield{author}{\bibinfo{person}{David~P. Roy}, \bibinfo{person}{S.~N.
  Trigg}, \bibinfo{person}{Roy Bhima}, \bibinfo{person}{Bruce~H. Brockett},
  \bibinfo{person}{Opha~Pauline Dube}, \bibinfo{person}{Philip~E. Frost},
  \bibinfo{person}{Navashni Govender}, \bibinfo{person}{Tobias Landmann},
  \bibinfo{person}{Johan~Le Roux}, \bibinfo{person}{Tsepo Lepono},
  \bibinfo{person}{Joaquim Macuacua}, \bibinfo{person}{Cheikh Mbow},
  \bibinfo{person}{K.~L. Mhwandangara}, \bibinfo{person}{Belda Mosepele},
  \bibinfo{person}{Onisimo Mutanga}, \bibinfo{person}{Gosiame Neo-Mahupeleng},
  \bibinfo{person}{Mario Norman}, {and} \bibinfo{person}{S. Virgilo}.}
  \bibinfo{year}{2006}\natexlab{}.
\newblock \showarticletitle{The utility of satellite fire product accuracy
  Information-perspectives and recommendations from the Southern Africa fire
  network}.
\newblock \bibinfo{journal}{\emph{IEEE Transactions on Geoscience and Remote
  Sensing}}  \bibinfo{volume}{44} (\bibinfo{year}{2006}),
  \bibinfo{pages}{1928--1930}.
\newblock


\bibitem[Singh et~al\mbox{.}(2021)]%
        {structurefireanalysis}
\bibfield{author}{\bibinfo{person}{Priya Singh}, \bibinfo{person}{Chandra
  Sabnani}, {and} \bibinfo{person}{Vijay Kapse}.}
  \bibinfo{year}{2021}\natexlab{}.
\newblock \showarticletitle{Hotspot Analysis of Structure Fires in Urban
  Agglomeration: A Case of Nagpur City, India}.
\newblock \bibinfo{journal}{\emph{Fire}}  \bibinfo{volume}{4}
  (\bibinfo{date}{07} \bibinfo{year}{2021}), \bibinfo{pages}{38}.
\newblock
\urldef\tempurl%
\url{https://doi.org/10.3390/fire4030038}
\showDOI{\tempurl}


\bibitem[Stefanidou et~al\mbox{.}(2008)]%
        {urbanfiresmoke}
\bibfield{author}{\bibinfo{person}{Maria Stefanidou}, \bibinfo{person}{Sotiris
  Athanaselis}, {and} \bibinfo{person}{Chara Spiliopoulou}.}
  \bibinfo{year}{2008}\natexlab{}.
\newblock \showarticletitle{Health Impacts of Fire Smoke Inhalation}.
\newblock \bibinfo{journal}{\emph{Inhalation toxicology}}  \bibinfo{volume}{20}
  (\bibinfo{date}{07} \bibinfo{year}{2008}), \bibinfo{pages}{761--6}.
\newblock
\urldef\tempurl%
\url{https://doi.org/10.1080/08958370801975311}
\showDOI{\tempurl}


\bibitem[Stephens et~al\mbox{.}(2001)]%
        {airnow}
\bibfield{author}{\bibinfo{person}{G. Stephens}, \bibinfo{person}{D. McNamara},
  \bibinfo{person}{R. Fennimore}, \bibinfo{person}{B. Ramsay}, {and}
  \bibinfo{person}{Mark Ruminski}.} \bibinfo{year}{2001}\natexlab{}.
\newblock \showarticletitle{NOAA's Improved Fire and Smoke Analysis, A Global
  Disaster Information Network Initiative}.
\newblock \bibinfo{journal}{\emph{AGU Spring Meeting Abstracts}}
  \bibinfo{volume}{-1} (\bibinfo{date}{04} \bibinfo{year}{2001}).
\newblock


\bibitem[Toon et~al\mbox{.}(2010)]%
        {smokeinhalation}
\bibfield{author}{\bibinfo{person}{Michael Toon}, \bibinfo{person}{Marc
  Maybauer}, \bibinfo{person}{John Greenwood}, \bibinfo{person}{Dirk Maybauer},
  {and} \bibinfo{person}{John Fraser}.} \bibinfo{year}{2010}\natexlab{}.
\newblock \showarticletitle{Management of acute smoke inhalation injury}.
\newblock \bibinfo{journal}{\emph{Critical care and resuscitation : journal of
  the Australasian Academy of Critical Care Medicine}}  \bibinfo{volume}{12}
  (\bibinfo{date}{03} \bibinfo{year}{2010}), \bibinfo{pages}{53--61}.
\newblock


\bibitem[(USA et~al\mbox{.}(2014)]%
        {inbook}
\bibfield{author}{\bibinfo{person}{Christopher (USA}, \bibinfo{person}{Vicente
  Barros}, \bibinfo{person}{Michael Mastrandrea}, \bibinfo{person}{Katharine
  (USA}, \bibinfo{person}{Mohamed Abdrabo}, \bibinfo{person}{W. Adger},
  \bibinfo{person}{Yury Federation}, \bibinfo{person}{O. Anisimov},
  \bibinfo{person}{Doug Arent}, \bibinfo{person}{Jonathon (Australia},
  \bibinfo{person}{Virginia (USA}, \bibinfo{person}{Rongshuo (China},
  \bibinfo{person}{Monalisa (USA/India}, \bibinfo{person}{Stewart Cohen},
  \bibinfo{person}{Wolfgang (Germany/France}, \bibinfo{person}{Purnamita
  Dasgupta}, \bibinfo{person}{Debra Davidson}, \bibinfo{person}{Fatima Denton},
  \bibinfo{person}{Petra Doell}, {and} \bibinfo{person}{Gary Yohe}.}
  \bibinfo{year}{2014}\natexlab{}.
\newblock \bibinfo{booktitle}{\emph{Climate change 2014: impacts, adaptation,
  and vulnerability – IPCC WGII AR5 summary for policymakers}}.
\newblock \bibinfo{pages}{1--32}.
\newblock


\bibitem[Vrabi et~al\mbox{.}(2018)]%
        {Vrabi2018DigitalTU}
\bibfield{author}{\bibinfo{person}{Rok Vrabi}, \bibinfo{person}{John~Ahmet
  Erkoyuncu}, \bibinfo{person}{Peter Butala}, {and} \bibinfo{person}{Rajkumar
  Roy}.} \bibinfo{year}{2018}\natexlab{}.
\newblock \showarticletitle{Digital twins: Understanding the added value of
  integrated models for through-life engineering services}.
\newblock \bibinfo{journal}{\emph{Procedia Manufacturing}}
  \bibinfo{volume}{16} (\bibinfo{year}{2018}), \bibinfo{pages}{139--146}.
\newblock


\bibitem[Wang et~al\mbox{.}(2010)]%
        {fluidsimprogress}
\bibfield{author}{\bibinfo{person}{S.-J Wang}, \bibinfo{person}{C. Xu},
  \bibinfo{person}{P. Yuan}, {and} \bibinfo{person}{Y.-Y Wang}.}
  \bibinfo{year}{2010}\natexlab{}.
\newblock \showarticletitle{Research progress of realistic fluid simulation at
  home and abroad}.
\newblock   \bibinfo{volume}{22} (\bibinfo{date}{02} \bibinfo{year}{2010}),
  \bibinfo{pages}{280--286}.
\newblock


\bibitem[Wu et~al\mbox{.}(2020)]%
        {covid}
\bibfield{author}{\bibinfo{person}{X. Wu}, \bibinfo{person}{R.~C. Nethery},
  \bibinfo{person}{M.~B. Sabath}, \bibinfo{person}{D. Braun}, {and}
  \bibinfo{person}{F. Dominici}.} \bibinfo{year}{2020}\natexlab{}.
\newblock \showarticletitle{Air pollution and covid-19 mortality in the United
  States: Strengths and limitations of an ecological regression analysis}.
\newblock \bibinfo{journal}{\emph{Science Advances}} \bibinfo{volume}{6},
  \bibinfo{number}{45} (\bibinfo{year}{2020}).
\newblock
\urldef\tempurl%
\url{https://doi.org/10.1126/sciadv.abd4049}
\showDOI{\tempurl}


\bibitem[Xie et~al\mbox{.}(2018)]%
        {10.1145/3197517.3201304}
\bibfield{author}{\bibinfo{person}{You Xie}, \bibinfo{person}{Erik Franz},
  \bibinfo{person}{Mengyu Chu}, {and} \bibinfo{person}{Nils Thuerey}.}
  \bibinfo{year}{2018}\natexlab{}.
\newblock \showarticletitle{TempoGAN: A Temporally Coherent, Volumetric GAN for
  Super-Resolution Fluid Flow}.
\newblock \bibinfo{journal}{\emph{ACM Trans. Graph.}} \bibinfo{volume}{37},
  \bibinfo{number}{4}, Article \bibinfo{articleno}{95} (\bibinfo{date}{jul}
  \bibinfo{year}{2018}), \bibinfo{numpages}{15}~pages.
\newblock
\showISSN{0730-0301}
\urldef\tempurl%
\url{https://doi.org/10.1145/3197517.3201304}
\showDOI{\tempurl}


\bibitem[Zhao et~al\mbox{.}(2022)]%
        {ZHAO2022101755}
\bibfield{author}{\bibinfo{person}{Jiaming Zhao}, \bibinfo{person}{Zhisheng
  Xu}, \bibinfo{person}{Houlin Ying}, \bibinfo{person}{Xueqi Guan},
  \bibinfo{person}{Kunkun Chu}, \bibinfo{person}{Sylvain~Marcial {Sakepa
  Tagne}}, {and} \bibinfo{person}{Haowen Tao}.}
  \bibinfo{year}{2022}\natexlab{}.
\newblock \showarticletitle{Study on smoke spread characteristic in urban
  interval tunnel fire}.
\newblock \bibinfo{journal}{\emph{Case Studies in Thermal Engineering}}
  \bibinfo{volume}{30} (\bibinfo{year}{2022}), \bibinfo{pages}{101755}.
\newblock
\showISSN{2214-157X}
\urldef\tempurl%
\url{https://doi.org/10.1016/j.csite.2022.101755}
\showDOI{\tempurl}


\end{thebibliography}

\appendix

\newpage 

\section{Online Resources}

To help others use our model, we have made all source code developed publicly available on GitHub, along with documentation. We have divided the work into the following three repositories.

\textbf{FireIncidentFrontend} 

\href{https://github.com/UrbanInfoLab/FireIncidentFrontend}{https://github.com/UrbanInfoLab/FireIncidentFrontend}

This repository contains strictly the front-end files required to run our model. The 2D and 3D front-ends are included, both being statically served via JavaScript libraries through MapBox and Leaflet. One can also find the code to display both the 2D (.KML) and 3D (.FBX) predicted smoke outputs in their browser with accurate positioning on a geographic map.

\textbf{FireIncidentBackend} 

\href{https://github.com/UrbanInfoLab/FireIncidentBackend}{https://github.com/UrbanInfoLab/FireIncidentBackend}

This repository contains strictly the back-end files required to run our model. The base back-end finds itself in a curated list of executable events to fetch, clean, and synthesize data on an hourly basis from chosen fire departments. Our custom implementation of the VSmoke algorithm is also included, alongside our MantaFlow algorithm, with proper usage examples.

\textbf{FireIncidentData} 

\href{https://github.com/UrbanInfoLab/FireIncidentData}{https://github.com/UrbanInfoLab/FireIncidentData}

This repository contains all accumulated data from running our model for the last year. Multiple cities, like Los Angeles, Houston, Seattle, and others are included in addition to Austin, and all data is collected and updated on a nightly basis to the repository. We also provide access to an API integrating with our live server in case the data is needed on a real-time basis.

\end{document}